\documentstyle[epsfig]{mn}

\newif\ifAMStwofonts

\newcommand{\tf}{t_{\rm fwhm}}

\newcommand{\tF}{t_F}
\newcommand{\tE}{t_E}
\newcommand{\Dfm}{\Delta F_{\rm max}}
\newcommand{\oF}{\omega_F}
\newcommand{\MJ}{M_{\rm Jup}}

\ifoldfss
  \ifCUPmtlplainloaded \else
    \NewTextAlphabet{textbfit} {cmbxti10} {}
    \NewTextAlphabet{textbfss} {cmssbx10} {}
    \NewMathAlphabet{mathbfit} {cmbxti10} {} 
    \NewMathAlphabet{mathbfss} {cmssbx10} {} 
  \fi
  \ifAMStwofonts
    \ifCUPmtlplainloaded \else
      \NewSymbolFont{upmath} {eurm10}
      \NewSymbolFont{AMSa} {msam10}
      \NewMathSymbol{\upi}     {0}{upmath}{19}
      \NewMathSymbol{\umu}     {0}{upmath}{16}
      \NewMathSymbol{\upartial}{0}{upmath}{40}
      \NewMathSymbol{\leqslant}{3}{AMSa}{36}
      \NewMathSymbol{\geqslant}{3}{AMSa}{3E}

    \fi
  \fi
\fi 

\ifnfssone
  \newmathalphabet{\mathit}
  \addtoversion{normal}{\mathit}{cmr}{m}{it}
  \addtoversion{bold}{\mathit}{cmr}{bx}{it}
  \newmathalphabet{\mathbfit} 
  \addtoversion{normal}{\mathbfit}{cmr}{bx}{it}
  \addtoversion{bold}{\mathbfit}{cmr}{bx}{it}
  \newmathalphabet{\mathbfss} 
  \addtoversion{normal}{\mathbfss}{cmss}{bx}{n}
  \addtoversion{bold}{\mathbfss}{cmss}{bx}{n}
  \ifAMStwofonts
    \ifCUPmtlplainloaded \else
      \UseAMStwoboldmath
      \makeatletter
      \new@mathgroup\upmath@group
      \define@mathgroup\mv@normal\upmath@group{eur}{m}{n}
      \define@mathgroup\mv@bold\upmath@group{eur}{b}{n}
      \edef\UPM{\hexnumber\upmath@group}
      \new@mathgroup\amsa@group
      \define@mathgroup\mv@normal\amsa@group{msa}{m}{n}
      \define@mathgroup\mv@bold\amsa@group{msa}{m}{n}
      \edef\AMSa{\hexnumber\amsa@group}
      \makeatother
      \mathchardef\upi="0\UPM19
      \mathchardef\umu="0\UPM16
      \mathchardef\upartial="0\UPM40
      \mathchardef\leqslant="3\AMSa36
      \mathchardef\geqslant="3\AMSa3E
    \fi
  \fi
\fi 

\ifnfsstwo
  \DeclareMathAlphabet{\mathbfit}{OT1}{cmr}{bx}{it}
  \SetMathAlphabet\mathbfit{bold}{OT1}{cmr}{bx}{it}
  \DeclareMathAlphabet{\mathbfss}{OT1}{cmss}{bx}{n}
  \SetMathAlphabet\mathbfss{bold}{OT1}{cmss}{bx}{n}
  \ifAMStwofonts
    \ifCUPmtlplainloaded \else
      \DeclareSymbolFont{UPM}{U}{eur}{m}{n}
      \SetSymbolFont{UPM}{bold}{U}{eur}{b}{n}
      \DeclareSymbolFont{AMSa}{U}{msa}{m}{n}
      \DeclareMathSymbol{\upi}{0}{UPM}{"19}
      \DeclareMathSymbol{\umu}{0}{UPM}{"16}
      \DeclareMathSymbol{\upartial}{0}{UPM}{"40}
      \DeclareMathSymbol{\leqslant}{3}{AMSa}{"36}
      \DeclareMathSymbol{\geqslant}{3}{AMSa}{"3E}
    \fi
  \fi
\fi 

\ifCUPmtlplainloaded \else
  \ifAMStwofonts \else 
    \def\upi{\pi}
    \def\umu{\mu}
    \def\upartial{\partial}
  \fi
\fi

\title{The Brown Dwarf Mass Function from Pixel Microlensing}
\author[E.A. Baltz and J. Silk]
	{E.A. Baltz$^1$ and J. Silk$^2$ \\
	$^1$ISCAP, Columbia Astrophysics Laboratory, 550 W 120th St.,
	Mail Code 5247,	New York, NY 10027, USA \\
	$^2$Nuclear and Astrophysics Laboratory, Keble Road, Oxford OX1 3RH}
\date{Accepted
      Received}

\pagerange{\pageref{firstpage}--\pageref{lastpage}}
\pubyear{2000}

\begin{document}
\maketitle

\label{firstpage}

\begin{abstract}
We argue that gravitational microlensing is a feasible technique for measuring
the mass function of brown dwarf stars in distant galaxies.  Microlensing
surveys of the bulge of M31, and of M87 in the Virgo cluster, may provide
enough events to differentiate the behaviour of the mass function of lenses
below the hydrogen burning limit (though we find that M87 is a more favourable
target).  Such objects may provide a significant supply of baryonic dark
matter, an interesting possibility for the study of galactic dynamics.
Furthermore, these systems have different metallicities than the solar
neighbourhood, which may affect the mass function.  These considerations are
relevant in the context of star formation studies.
\end{abstract}


\section{introduction}
Gravitational microlensing, by which dark objects are detected by their
magnification of bright sources being monitored, is a rapidly growing field,
with many applications in astrophysics.  Using microlensing to find dark
objects in the halo of the Milky Way was first proposed by Paczy\'nski (1986),
and elaborated by Griest (1991).  Several groups have extended the
possibilities of microlensing by the so-called pixel technique (Crotts 1992;
Baillon et al. 1993), where source stars are not resolved due to the high
degree of crowding.  However, successive images can be analysed, usually by
taking image differences, to uncover the brightenings due to variability in the
individual sources, some of which will be due to microlensing.  This technique
has been applied successfully by several collaborations (Tomaney \& Crotts
1996; Alcock et al. 1999a,b; Wo\'zniak 2000).  Previously, it has been
difficult to extract physical information from these pixel microlensing
lightcurves because of a degeneracy in the lightcurve shape with magnification.
The method of Gondolo (1999) for measuring the optical depth in pixel
microlensing without knowledge of the magnification in an event has been
extended to measurements of the mass function of lenses (Baltz 2000).  In this
paper we show how to apply this technique to measurements of the low-mass end
of the stellar mass function, as has been done in classical microlensing
surveys for brown dwarfs (Zhao, Spergel \& Rich 1995) and planetary-mass
objects (Alcock et al. 1998) in our own galaxy.  We hope to extend such limits
to distant galaxies as well.

\section{brown dwarfs}

Brown dwarfs are defined to lie in the mass range between the hydrogen-burning
limit (75 $\MJ$) and the deuterium burning threshold (13 $\MJ$: $1 M_{\odot} =
1047 \MJ)$.  A theoretical review of these objects is given by Chabrier and
Baraffe (2000).  Radial velocity searches have detected both brown dwarfs and
giant planets from $75 \MJ$ to $0.25 \MJ$ in orbits around solar-type stars.
The observational situation for brown dwarfs is reviewed by Basri (2000).
Studies of young star-forming regions have recently found evidence for isolated
objects in the giant planet mass range ($5-15 \MJ)$ (Zapateo-Ossorio et
al. 2000).  Isolated brown dwarfs have also been reported in various
star-forming regions (Lucas \& Roche 2000; Hillenbrand \& Carpenter 2000;
Najita, Tiedo \& Carr 2000).  The frequency of such objects is not known, but
they seem to be sufficiently numerous that the IMF must certainly continue from
the hydrogen burning limit to the giant planet regime, and could either be
flat, exhibit a weak power law decline, or even rise below the brown dwarf
limit.

Observations of isolated brown dwarfs and giant planets are necessarily
focussed on searching in star-forming regions, via the technique of infrared
imaging.  We show here that gravitational microlensing provides a potentially
powerful technique for exploring the initial mass function of brown dwarfs in
distant galaxies.

We investigate several possible mass functions for this type of object, below
the hydrogen burning limit.  Our proposed method will not be able to
distinguish fine features in the mass function, so we investigate several toy
mass functions.  It is the gross features of the mass function that we are
immediately concerned with, and simple model mass functions will suffice.  We
consider the class of mass functions (taking $M$ in units of $M_\odot$, and
with arbitrary normalisation)
\begin{equation}
\log\left(\frac{dN}{d\log M}\right)=
-\left[A\,\log M+B\left(1-\log M\right)^2\right],
\end{equation}
where $A$ and $B$ are free parameters (note that $B=0$ corresponds to a pure
power--law mass function $dN/d\,\log M\propto M^{-A}$).  These mass functions
agree at ten solar masses, and taking $A=1.35$, have the Salpeter (1955) slope
at ten solar masses.  This prescription ensures that the mass functions produce
equal light, as most of the visible light from stellar populations comes from
massive stars, heavier than the sun.  The uncertainty in the population thus
lies in the mass to light ratio, not in the total light, which is as it should
be.

Taking $B=0.25$ gives a mass function with approximately the Salpeter (1955)
slope down to $0.02 M_\odot$.  This model has a large amount of stellar mass
below the hydrogen burning limit, indicating a large amount of baryonic dark
matter locked up in brown dwarf stars.  Taking $B=0.32$ gives a mass function
turning over at around $0.08 M_\odot$, similar to the Miller--Scalo (1979) mass
function.  Lastly, we take $B=0.48$, giving a mass function turning over at
around $0.4 M_\odot$, similar to that of Gould, Bahcall \& Flynn (1996).  This
model has the least amount of baryonic dark matter.

\section{surveys of distant galaxies}
Pixel microlensing surveys of large distant galaxies are quite attractive for
learning about both the dynamical properties of the systems and the populations
of objects from which they are made.  We will discuss two possible targets to
illustrate the capabilities of this technique with ground and space based
telescopes.

A pixel microlensing survey of M31, the Andromeda Galaxy, is quite feasible
using a three-meter class ground based telescope.  This large spiral galaxy is
about 725 kpc distant, and is the nearest large galaxy to the Milky Way.  We
are primarily interested in the bulge of this galaxy, as it affords the densest
star fields and the largest optical depth for star--star lensing.  However,
finite source size effects for this target would require nearly continuous
monitoring for reasonable sensitivity at low masses to be achieved, so we will
not consider M31 further.

Using a space-based telescope, the reach of the pixel technique is much longer.
The giant elliptical galaxy M87, at the center of the Virgo cluster at a
distance of about 15 Mpc, is well within the reach of a pixel microlensing
survey by the Hubble Space Telescope (HST).  We take a mass model based on the
work of Tsai (1993),
\begin{equation}
\rho(r)=3.8\left[1+\left(\frac{r}{{\rm kpc}}\right)^2\right]^{-\alpha}\,
M_\odot\;{\rm pc}^{-3},
\end{equation}
\begin{equation}
\alpha=\max\left[1,
1+0.275\,\log\left(\frac{r}{\rm kpc}\right)\right].
\end{equation}

\begin{figure}
\epsfig{file=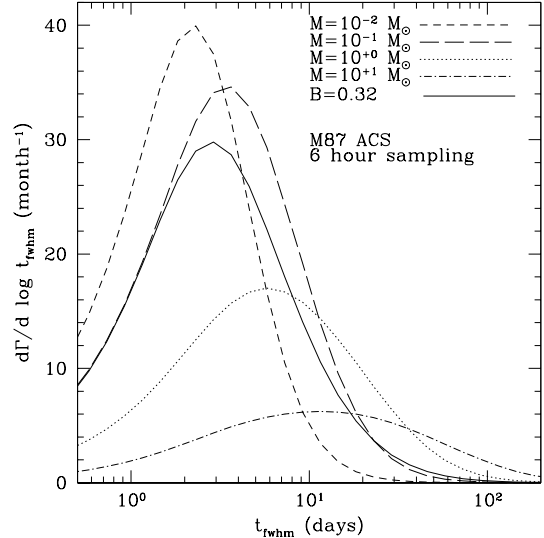,width=3in}
\caption{Self-lensing event rate towards M87 bulge with ACS on HST.  The sample
spacing is four times daily.  There is a much lower sensitivity to lenses below
about $10^{-2}\;M_\odot$.  We illustrate delta function mass functions,
together with the $B=0.32$ Miller-Scalo type mass function.  }
\label{acsrate}
\end{figure}

\begin{figure}
\epsfig{file=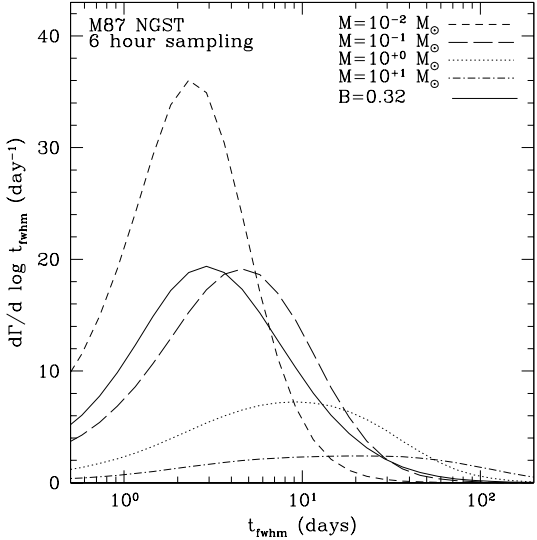,width=3in}
\caption{Self-lensing event rate towards M87 bulge with NGST.  The sample
spacing is four times daily.  The sensitivity drops considerably below about
$10^{-3}\;M_\odot$.  The mass functions are the same as in
Fig.~\protect\ref{acsrate}.}
\label{ngstrate}
\end{figure}

The Advanced Camera for Surveys (ACS) is an ideal instrument for this type of
observation.  The future NGST, planned as an eight meter class space telescope,
will allow a much more thorough survey, with better statistics.  For both
telescopes we assume exposures of thirty minutes (for the HST ACS, this is
possible in one orbit), taken every six hours.  More frequent sampling is
punishing in that fewer events are detected with the same telescope resources
(say 30 orbits for HST), though smaller masses can be probed.  Less frequent
sampling may allow more events to be detected with the same telescope
resources, but at the cost of the low-mass sensitivity that we desire for this
program.  For our purposes, six-hour sampling is about ideal, though with the
NGST there is more room to take a larger sample spacing.  We show the event
rates for these two telescopes below, in Figs.~\ref{acsrate}~and~\ref{ngstrate}
respectively (Baltz \& Silk 2000).  Rates for the NGST assume a field of view
4' square, with 50\% more throughput than the HST ACS, and nine times the
collection area.  In all cases in this paper, we assume that seven samples
above $2\sigma$ must be collected, as Criteria A of Alcock et al. (2000).

\section{measuring the mass function}
For each pixel microlensing event, we make in effect two measurements.  These
are the flux increase at maximum $\Dfm$, and the full-width at half maximum
time, $\tf$.  As Wo\'zniak \& Paczy\'nski (1997) have clearly shown, even in
the classical microlensing case, the shape of the lightcurve does not give much
more information than these two parameters.  However, in the classical
microlensing case, an additional measurement, that of the unlensed flux of the
source star, is also made, allowing an accurate determination of the
magnification of the event.

The Einstein times of events have been used to estimate the optical depth to
microlensing, and also to estimate the masses of the lenses.  However, to make
a good estimate of the Einstein time from a microlensing event, both $\tf$ and
the magnification are required.  We thus seek a different characteristic of
events.  The time-scale $\tf$ is problematic, as it strongly depends on the
unknown magnification.  We instead follow Gondolo (1999) and use the quantity
$\oF=\Dfm\tf$, which is effectively the Einstein time multiplied by the flux of
the source star.  This quantity is easily measured from pixel microlensing
events.  We choose to further normalize by dividing by the surface brightness
fluctuation flux ($\overline{F}\propto 10^{-0.4\overline{m}}$),
\begin{equation}
\tF=\frac{\Dfm}{\overline{F}}\frac{\tf}{\sqrt{3}}.
\label{eq:tF}
\end{equation}
The SBF flux $\overline{F}$ is measured by characterizing the pixel-to-pixel
variations of the surface brightness of a galaxy image (Tonry \& Schneider
1988).  This normalisation gives the approximate relation, with $\beta$ being
the event's minimum impact parameter in Einstein units,
\begin{equation}
\tF=\tE10^{-0.4(m-\overline{m})}\left(1-\frac{7}{3}\,\beta+
\frac{311}{72}\,\beta^2+O\left(\beta^3\right)\right)
\end{equation}
when $\beta\ll 1$, or equivalently, at high magnification.  Gondolo (1999)
showed how to measure the optical depth to microlensing using this time-scale.
We will use an extension of that method, described in more detail in Baltz
(2000), to constrain the mass function of the lenses, in this case brown
dwarfs.  The method is based on producing a weighted histogram of event rate
with flux time-scale $\tF$.  Differentially, we want to study the dimensionless
rate
\begin{equation}
N_F=\tF\frac{d\Gamma}{d\log\tF},
\label{eq:dimrate}
\end{equation}
given a delta-function mass function.  The quantity $N_F$ then encodes the
response of a microlensing survey to lenses of a given mass, effectively the
number of events expected per decade in time-scale when monitoring for a time
equal to the event time-scale.  Equivalently, this quantity is closely related
to the optical depth contributed by lenses producing events at a specific
time-scale.

For comparison, we can form quantities analogous to $N_F$ using the other
relevant time-scales, namely the Einstein time ($N_E$) and the full-width at
half maximum time ($N_{\rm fwhm}$).  Their definitions exactly mimic
Eq.~\ref{eq:dimrate}.

\begin{figure}
\epsfig{file=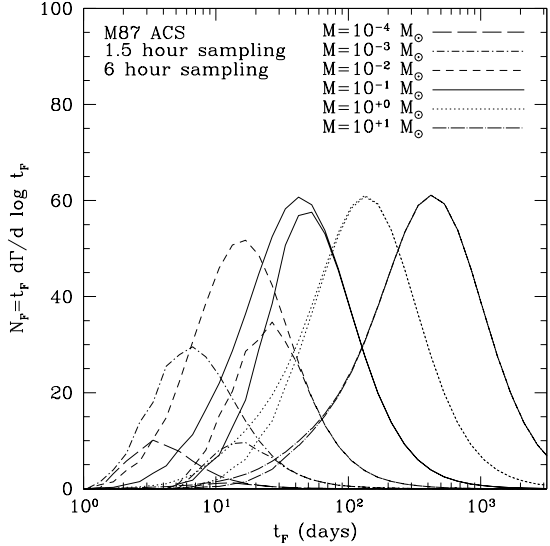,width=3in}
\caption{Dimensionless rate $N_F$ for ACS on HST.  The curves are, left to
right, for lenses of a single mass: $10^{-4,-3,-2,-1,0,1}M_\odot$.  Samples are
taken every 1.5 hours in the top curves, with the lower curves sampling every
six hours.  With six hour sampling, the sensitivity begins to decline at masses
less than $10^{-1}M_\odot$, and is quite low below $10^{-2}M_\odot$.}
\label{acscurves}
\end{figure}

\begin{figure}
\epsfig{file=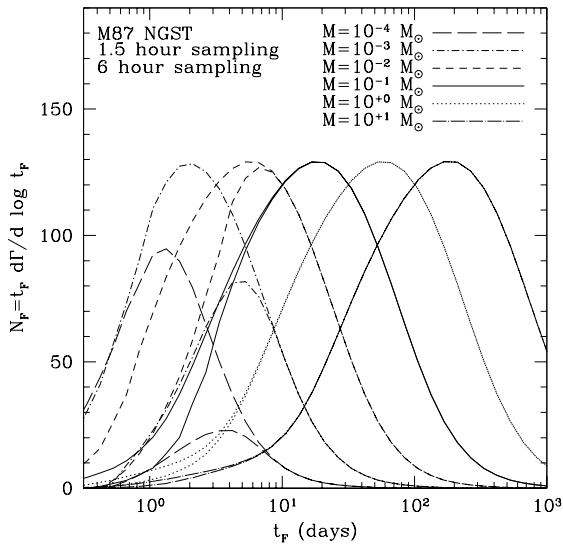,width=3in}
\caption{Dimensionless rate $N_F$ for NGST.  The curves are analogous to those
in Fig.~\ref{acscurves}.  With six hour sampling, the sensitivity only begins
to seriously decline at masses less than $10^{-2}M_\odot$.}
\label{ngstcurves}
\end{figure}

\begin{figure}
\epsfig{file=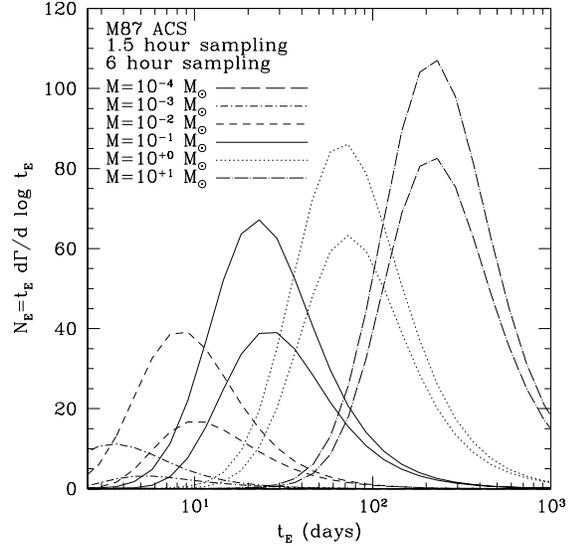,width=3in}
\caption{Dimensionless rate $N_E$ for ACS on HST.  The curves are analogous to
those in Fig.~\ref{acscurves}, and in fact represent the same microlensing
survey, with the time-scale $\tE$ rather then $\tF$.}
\label{acstEcurves}
\end{figure}

At fixed mass, one would naively think that the Einstein time is most useful,
as the distribution is the narrowest of the three.  However, the Einstein time
can be measured only rarely.  Furthermore, we find that the shape of the rate
distribution with the Einstein time is sensitive to the exact details of the
cuts used to define events.  Interestingly, the shape of the rate distribution
with the flux time is much less sensitive to the exact definition of an event,
which is clearly a desirable feature.  This happens because changing the
definition of an event usually amounts to changing the minimum value of the
flux increase at maximum.  This will simply change the short time-scale end of
the $\tF$ distribution, leaving the peak shape intact in most cases.  This
point is further explored in Baltz (2000).  Thus, it seems that the time-scale
$\tF$ is perhaps preferable to the Einstein time in extracting the parameters
of the source--lens system.  In Figs.~\ref{acscurves}~and~\ref{ngstcurves}, we
display the function $N_F$, given delta-function mass functions.  Effectively,
these are the smoothing kernels over which the true mass function can be
measured.  In Fig.~\ref{acstEcurves} we plot the function $N_E$, based on the
Einstein time-scale, for the same exact parameters as for the ACS in
Fig.~\ref{acscurves}.  As is clearly evident, the function $N_E$ is more
sensitive {\em near its peak} than $N_F$ to changing the effective threshold,
as is the case for changing the sampling strategy.  Thus the discussion in this
paragraph is validated: even if $\tE$, and thus $N_E$, were measurable, $N_F$
is preferred in cases where there is a high sensitivity to the threshold for
events.

The universality of the shape is broken by two effects: finite source size and
finite time sampling.  The primary effect of the finite source size is that the
magnification is bounded by the fact that the source stars are not point
sources.  This effect is obviously more pronounced for smaller Einstein rings,
and thus for smaller masses.  A finite time sampling implies a minimum event
time-scale that can be detected.  Lower mass lenses of course produce shorter
events, thus more of the low-mass events are missed.  The finite total
observing time would also subtract very long events, but for our purposes, this
is not a concern, as we are interested in the short events due to 0.01-0.1
$M_\odot$ stars.

\section{discussion}

\begin{figure}
\epsfig{file=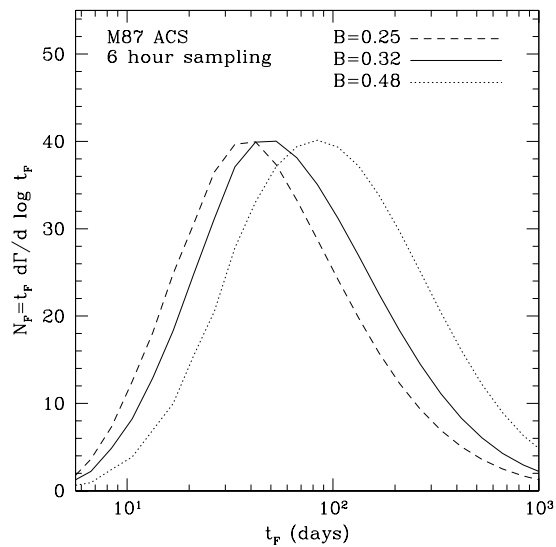,width=3in}
\caption{Mass functions with ACS on HST.  We illustrate the measured $N_F$ for
the three basic mass functions.  The normalisations of the $B=0.25$ and
$B=0.48$ curves have been adjusted to compare the peak positions with the
$B=0.32$ case.}
\label{acsmf}
\end{figure}

\begin{figure}
\epsfig{file=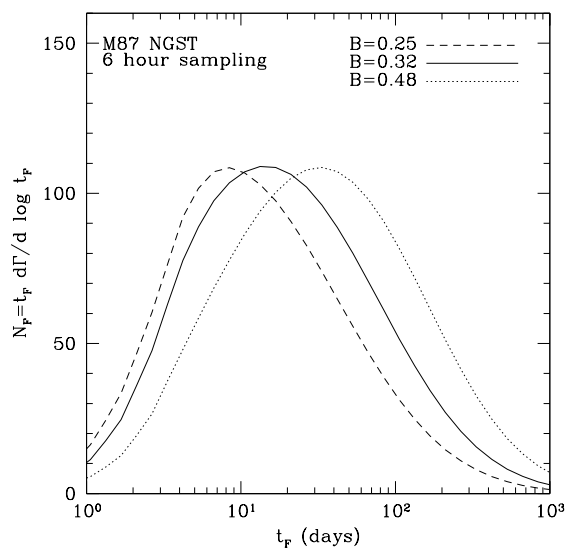,width=3in}
\caption{Mass functions with NGST.  We illustrate the measured $N_F$ for the
three basic mass functions.  The normalisations have been adjusted as in
Fig.\ref{acsmf}}
\label{ngstmf}
\end{figure}

We now illustrate the full $N_F$ distribution, integrated over the simple mass
function discussed in Sec.~2.  As discussed previously, $N_F$ for a single mass
acts as a smoothing filter for the mass function.  We show the results for the
HST ACS in Fig.~\ref{acsmf} and the NGST in Fig.~\ref{ngstmf}.  The pure
Salpeter mass function has a peak at the shortest time-scale, the flattened
models (Miller-Scalo and Salpeter with cutoff) has a longer peak, and the
GBF-type model has the longest peak time-scale.  With these surveys, the
statistics should be sufficient to distinguish these models. We should state
that the normalisation alone is probably insufficient to distinguish models,
but the position in time-scale of the peak of the rate distribution is quite
robust.  In fact, we have adjusted the normalisation of the $B=0.3$ and
$B=0.75$ models so that the peak rates agree, in order to compare the
time-scales.  For the surveys we consider, the peak time-scale differs by a
factor of 2-3 between a mass function that is flat for substellar masses
(Miller-Scalo), and one that that is sharply declining at those masses (GBF).
We note that mass functions that differ only below 0.1 $M_\odot$ will be quite
difficult to distinguish, but for mass functions which begin to differ at
around a solar mass, the microlensing technique is quite powerful.

We have shown that pixel microlensing can be a powerful tool for measuring the
mass function of low mass and brown dwarf stars, less massive than the sun.
Since this technique is effective to very large distances, we have a chance to
learn something about the universal properties of brown dwarf mass functions.

\section*{acknowledgments}

We thank P. Gondolo, E. Kerins, N.W. Evans and the anonymous referee for useful
conversations.

\bsp

\label{lastpage}

\end{document}